\RequirePackage{amsthm}

\documentclass[sn-mathphys,Numbered]{sn-jnl}% Math and Physical Sciences Reference Style
\newcommand{\norm}[1]{\left\lVert#1\right\rVert}
\renewcommand{\Im}{\operatorname{Im}}
\renewcommand{\Re}{\operatorname{Re}}

\usepackage{graphicx}%
\usepackage{multirow}%
\usepackage{amsmath,amssymb,amsfonts}%
\usepackage{amsthm}%
\usepackage{mathrsfs}%
\usepackage[title]{appendix}%
\usepackage{xcolor}%
\usepackage{textcomp}%
\usepackage{manyfoot}%
\usepackage{booktabs}%
\usepackage{algorithm}%
\usepackage{algorithmicx}%
\usepackage{algpseudocode}%
\usepackage{listings}%

\DeclareMathOperator{\Tr}{Tr}

\theoremstyle{thmstyleone}%
\theoremstyle{thmstyletwo}%

\theoremstyle{thmstylethree}%
\newtheorem*{nntheorem}{Theorem}

\begin{document}

\title{Unitary time evolution in quantum mechanics is a stronger physical postulate than linear time evolution}


\author*{\fnm{Edward} \sur{Parker}}\email{tparker@alumni.physics.ucsb.edu}

\affil{
\orgaddress{\city{Washington}, \state{DC}, \country{USA}}}


\abstract{Discussions of quantum mechanics often loosely claim that time evolution logically must be unitary, in order for the probabilistic interpretation of the amplitudes of the state vector to make sense at all times. We discuss from first principles whether this claim is true: if we assume only that the time-evolution operator is \emph{linear}, then does the stronger requirement that it be \emph{unitary} follow from the other axioms of quantum mechanics? The answer is subtle. We discuss two mathematically distinct but physically equivalent formulations of the axioms of quantum mechanics, and consider generalizing each to postulate only that time evolution is linear. Within one formulation, the unitarity of time evolution follows logically from the other axioms -- but within the other formulation, it does not. Allowing the time-evolution operator to be (a priori) arbitrarily linear does not change the physical observables in one formulation of quantum mechanics, but changes the other formulation to a \emph{distinct} (internally consistent) physical theory that allows new phenomenology like (e.g.) faster-than-light communication. Therefore, the unitarity of time evolution is arguably better thought of as a logically independent and experimentally falsifiable axiom of quantum mechanics, not as a tautological consequence of the other axioms.}

\keywords{Unitarity, time evolution, quantum mechanics, normalization}

\maketitle

\section{Introduction}

Discussions of quantum mechanics (QM) often claim that time evolution logically must be unitary in order to preserve total probability \cite{ACM, MIT, Littlejohn}. A more precise version of this claim states that the basic rules of probability require that the norm of a quantum state vector must be preserved over time -- so if we postulate only that time evolution is represented by a \emph{linear} operator $\hat{U}$, then it must in fact be unitary, because unitary operations are the only linear operators on an inner product space that preserve vector norms.\footnote{Throughout this paper, we assume the standard norm on an inner product space $\norm{\psi} := \sqrt{\langle \psi | \psi \rangle}$. Ref.~\cite{Aaronson} considers other choices of norm for vectors in $\mathbb{C}^n$ and finds that only the standard inner product space $2$-norm (and the $1$-norm of classical probability theory, if we restrict ourselves to nonnegative entries) permit nontrivial norm-preserving linear maps.}

This article attempts to clarify certain implicit assumptions behind this claim. We argue that there are two physically equivalent ways to formulate ``textbook'' quantum mechanics with unitary time evolution. Under one formulation, the unitarity of time evolution does follow naturally from the other postulates of QM and the assumption of linear time-evolution. But the other formulation permits a natural generalization of the time-evolution rule to allow non-unitary time evolution that is fully compatible with the other postulates, but which is not allowed by textbook QM. Moreover, this generalized time-evolution rule is not just a different mathematical formalism for standard QM, but represents a genuinely different physical theory that allows for physical phenonema that are impossible under the axioms of standard QM. Therefore, under the second formulation of QM, the question whether time evolution is unitary is directly experimentally testable.

(Of course, all real-world experimental evidence so far is fully compatible with the proposition that time evolution is indeed unitary. In this paper, we are not proposing an alternative to standard QM as an actual theory of the real world; our central argument is simply that the unitarity of time evolution does indeed need to be specified as an independent axiom of standard QM, which -- at least under certain formulations of QM -- is logically independent of the other axioms.)

The non-standard alternative version of QM mentioned above has been discussed before (in a somewhat different context) in Ref.~\cite{Aaronson} under the name of ``manual normalization.'' Our goal is not to discuss its phenomenology in detail, but simply to clarify the conceptual points that (a) this theory allows for non-unitary time evolution even in the absence of any measurements, but (b) it is fully compatible with (certain formulations of) all of the axioms of standard QM \emph{other} than the requirement of unitary time evolution. Therefore, this theory serves as a concrete counterexample to the claim that the other axioms automatically imply that time evolution must be unitary.

We make the important caveat that we only consider the abstract version of QM that is often used in (e.g.) quantum information science, which allows for arbitrary unitary time evolution and does not explicitly postulate the Schr\"{o}dinger equation \cite{Mermin}. The unitarity of time evolution does indeed follow automatically from the Schr\"{o}dinger equation. But it does not (necessarily) follow automatically from the \emph{other} axioms of QM (like the Born rule), as is sometimes claimed.

This paper assumes familiarity with the basic axioms of quantum mechanics, but no other background (except during one short optional discussion). The footnotes go into a bit more mathematical detail than the main text does, but are not necessary for conveying the main argument.

\section{Two physically equivalent formulations of the axioms of quantum mechanics} \label{Formulations}

We will not attempt to be completely rigorous in our statement of the axioms of QM; in the usual tradition of physics, we will be just rigorous enough to get our point across, but no more.

We will follow the general axiom set laid out by Shankar \cite{Shankar}, omitting details that will not be necessary for our argument. We will deliberately present them somewhat vaguely at first, and then refine them with more details below.

In order to avoid the many mathematical subtleties that arise from infinite-dimensional Hilbert spaces, we will assume that all Hilbert spaces are finite-dimensional.

Moreover, in the main body of this article we will mostly follow Shankar's approach and focus on the simple version of QM that Ref.~\cite{Aaronson} whimsically refers to as ``QM lite''. In ``QM lite'', we only consider pure states and do not consider any mixed states or reduced density operators.\footnote{If you do not know what these terms mean, that is okay. If you understand the four axioms listed directly below, then most of this article should be accessible to you (except for the appendix).} In particular, we assume that the system of interest is isolated, so it remains pure at all times. We also do not consider the possibility of superselection rules. Accordingly, in this article we (mostly) do not consider multipartite systems, and we (mostly) ignore any tensor product structure that the Hilbert space may have. The only exception in the main body of this article is a single paragraph in Section~\ref{Generalizations}, where we consider a bipartite system in order to illustrate our results with a concrete example. In the appendix, we briefly touch on how the topics discussed in this paper generalize to the more complex cases of mixed states and multipartite systems; the appendix assumes somewhat more background in quantum information theory than the main body of this article does.

At a high level, the standard theory of QM -- neglecting the subtleties mentioned above -- can be derived from four basic axioms (with some clarifying details in the footnotes):

\begin{enumerate}
\item The state of an isolated quantum system is (non-uniquely) represented by a vector $|\psi\rangle$ in a complex Hilbert space.\footnote{``Non-uniquely represented'' means that multiple state vectors in the Hilbert space correspond to the same physical state.}
\item Physical observables are represented by Hermitian operators on the Hilbert space.\footnote{Some treatments, such as Shankar's, include the canonical commutation relation $\big[ \hat{X}_i, \hat{P}_j \big] = i \hbar\, \delta_{ij}\, \hat{I}$ as part of this axiom. But this commutation relation cannot be represented on the finite-dimensional Hilbert spaces that we are considering in this article, and we will not use it here.}
\item The rules for measurement: the possible outcomes of a measurement of a physical observable are the eigenvalues of the corresponding Hermitian operator $\hat{A}$. If a system is in state $|\psi\rangle$ immediately before the measurement is performed, then the probability of observing each eigenvalue $\lambda$ is proportional to $|\langle \lambda | \psi \rangle|^2$, where $|\lambda\rangle$ is an eigenvector of $\hat{A}$ with eigenvalue $\lambda$. Immediately after the measurement is performed, the system's state is the eigenvector $|\lambda\rangle$ corresponding to the observed eigenvalue.\footnote{If $\hat{A}$ is degenerate, then we instead use $P(\lambda) = \langle \psi | \hat{P}_\lambda | \psi \rangle$, where $\hat{P}_\lambda$ is the orthogonal projection operator onto the eigenspace of $\hat{A}$ corresponding to the eigenvalue $\lambda$. We ignore any interpretational questions about whether the measurement physically changes the system's ontic state or only updates the experimenter's epistemic description of the system.}
\item Absent an intervening measurement, time evolution from time $t_i$ to time $t_f$ is given by a unitary operator $\hat{U}(t_f, t_i)$. That is, a state $|\psi_i\rangle$ at time $t_i$ gets mapped to the state $|\psi_f\rangle = \hat{U}(t_f, t_i) |\psi_i\rangle$ at time $t_f$.\footnote{A linear operator $\hat{U}$ is unitary iff $\hat{U}^\dagger \hat{U} = \hat{I} = \hat{U} \hat{U}^\dagger$. For a finite-dimensional Hilbert space, either of those equalities automatically implies the other. The initial time $t_i$ and final time $t_f$ are often left implicit, and the term ``the time-evolution operator'' is used to refer to the entire family of unitary operators $\hat{U}(t_f,t_i)$. If the time-evolution operator is time-translationally invariant, then the family of operators $\hat{U}(\Delta t)$ parameterized by the time interval $\Delta t := t_f - t_i$ forms a one-parameter Lie group.}
\end{enumerate}

We note that axiom \#3, the rules for measurement, can be separated into two logically independent sub-axioms. The first sub-axiom is the Born rule for the probabilities of measuring each eigenvalue $\lambda$: $P(\lambda) \propto |\langle \lambda | \psi \rangle|^2$. The second sub-axiom is the ``update rule'': that after the measurement, the state of the system changes to an eigenstate of the measured observable. In this article, we will be almost entirely concerned with the first sub-axiom, the Born rule, and we will not discuss the update rule. We will think of time evolution as a single dynamical transition across some fixed time interval from an initial state to a final state -- potentially immediately followed by a single measurement -- and we will not consider more complicated issues such as repeated measurements or infinitesimally short time evolution.

The axiom set given above is not quite precise enough to be operationally useful. For our purposes, we will need to consider two physically equivalent -- but slightly formally different -- variant formulations A and B.

\subsection{Formulation A}

Variant A is the formulation that is perhaps more often taught in a first introduction to QM, since it is more convenient for concrete calculations:

\begin{enumerate}
\item[1A.] The state of an isolated quantum system is (non-uniquely) represented by a normalized \emph{unit} vector $|\psi\rangle$ in a complex Hilbert space.
\item[2A.] [Same as \#2.]
\item[3A.] [Same as \#3, except:] the probability $P(\lambda)$ of observing each eigenvalue $\lambda$ is
\[
P(\lambda) = |\langle \lambda | \psi \rangle|^2,
\]
where $|\lambda\rangle$ is a \emph{unit} eigenvector of $\hat{A}$.
\item[4A.] [Same as \#4.]
\end{enumerate}

In this formulation, the physical state of a quantum system is only specified up to an arbitrary complex \emph{phase factor} $e^{i \theta},\ \theta \in [0, 2\pi)$; the state vectors $|\psi\rangle$ and $e^{i \theta} |\psi\rangle$ correspond to the same physical quantum state. Put another way: the physical state of the system is \emph{uniquely} represented by an \emph{equivalence class} of unit state vectors with respect to the equivalence relation
\[
\left( |\psi\rangle \sim |\phi\rangle \right) \text{ iff } \left( \exists\, \theta \in [0, 2 \pi) \text{ such that } |\psi\rangle = e^{i \theta} |\phi\rangle \right).
\]

\subsection{Formulation B}

Variant B is sometimes used in more theoretical contexts:
\begin{enumerate}
\item[1B.] The state of an isolated quantum system is (non-uniquely) represented by a \emph{nonzero} vector $|\psi\rangle$ in a complex Hilbert space.
\item[2B.] [Same as \#2.]
\item[3B.] [Same as \#3, except:] the probability $P(\lambda)$ of observing each eigenvalue $\lambda$ is
\[
P(\lambda) = \frac{\langle \psi | \lambda \rangle \langle \lambda | \psi \rangle}{\langle \psi | \psi \rangle \langle \lambda | \lambda \rangle}.
\]
\item[4B.] [Same as \#4.]
\end{enumerate}

In this formulation, the physical state of a quantum system is only specified up to an arbitrary \emph{nonzero} complex number $z$; the state vectors $|\psi\rangle$ and $z |\psi\rangle$ correspond to the same physical quantum state. Put another way: the physical state of the system is \emph{uniquely} represented by an equivalence class of state vectors with respect to the equivalence relation
\[
\left( |\psi\rangle \sim |\phi\rangle \right) \text{ iff } \left( \exists\, z \in (\mathbb{C}\setminus \{0\}) \text{ such that } |\psi\rangle = z |\phi\rangle \right).
\]
For a given Hilbert space, the set of these equivalence classes is known as the corresponding \emph{projective Hilbert space}.\footnote{Confusingly, the elements of a projective Hilbert space are sometimes called \emph{points} and sometimes \emph{rays} -- although either the term ``line'' or (in the complex case) ``plane'' might arguably be a better analogy, since each element of a projective Hilbert space is a one-dimensional subspace of the original Hilbert space that is isomorphic to the underlying field. Also confusingly, a projective Hilbert space is not itself a Hilbert space, or even a vector space; there is no way to add together different elements of a projective Hilbert space.} 

The equivalence classes described above are somewhat abstract and unintuitive. But they have the advantage that each possible physical state of a quantum system corresponds to a \emph{unique} equivalence class, and the uniqueness of these representations is useful in many advanced theoretical applications. The projective Hilbert space formulation turns out to usually be more mathematically convenient than the equivalence classes defined within formulation A. So while formulation B may at first seem needlessly complicated, it is often used in mathematical physics \cite{Bengtsson}.

\subsection{Equivalence of formulations A and B} \label{equiv}

Formulations A and B are completely physically equivalent: the only difference is whether things get normalized before or after the inner product in the Born rule is taken. In formulation A, the state vector itself (and the observable eigenvectors) are normalized before any inner products are formed, so that the (norm-squared) inner products are already correctly normalized to sum to $1$ and represent direct probabilities. In formulation B, the (norm-squared) inner products are what get normalized to actual probabilities. In particular, both formulations yield the same probabilities $P(\lambda)$, which are the only quantities in QM that are physically measurable. These $P(\lambda)$ are guaranteed to lie in $[0, 1]$ and to add up to 1 when summed over all possible observable values $\lambda$, as must be the case by the definition of probability. Therefore, for standard QM, which formulation is more useful is largely a matter of taste (as well as a few practical or conceptual concerns mentioned above). But, as we discuss below, the two formulations admit physically distinct natural \emph{generalizations}.

Of course, we cannot necessarily consistently mix together axioms between the two formulations. If we were to combine together axioms 1B and 3A, then we would get nonsensical ``probabilities'' that do not sum to $1$ as required. But it turns out that we actually can consistently mix together axioms 1A and 3B, because if axiom 1A holds, then axiom 3B becomes equivalent to axiom 3A. The next section discusses in more detail which of the axioms above do or do not logically imply which (under which additional assumptions).

\section{Logical implications between axioms and possible generalizations} \label{Generalizations}

\begin{nntheorem}
If we assume axioms 1A-3A, but we weaken axiom 4A to only postulate that:
\begin{enumerate}
\item[4A'.] Time evolution is given by a \emph{linear} (but not a priori unitary) operator $\hat{U}(t_f, t_i)$,
\end{enumerate}
then axioms 1A-3A and 4A' still imply that $\hat{U}$ must be unitary.
\end{nntheorem}

This theorem is the rigorous version of the heuristic claim that ``conservation of probability requires that time evolution be unitary.''

\begin{proof}
Postulate 1A says that only \emph{unit} vectors $|\psi\rangle$ are legitimate state vectors (in formulation A), which is necessary in order for rule 3A to always yield valid probabilities that sum to $1$. So any time-evolution operator $\hat{U}$ must necessarily preserve the norm of all unit vectors: if $\norm{\psi} = 1$, then $\norm{\hat{U} | \psi \rangle} = \sqrt{\langle \psi | \hat{U}^\dagger \hat{U} | \psi \rangle}$ must equal $1$ as well.\footnote{We assume on physical grounds that time evolution must map a valid physical state to a valid physical state; states cannot ``self-destruct'' in finite time.

We are making an additional implicit assumption here. The argument presented here only applies to \emph{physical} states. Postulate 1A only says that all physical states are represented by unit vectors in the Hilbert space. Here, we are implictly assuming that the converse holds as well: that all unit vectors in the Hilbert space represent a valid physical state. Strictly speaking, we only prove here that $\hat{U}$ must preserve the norm of all \emph{physical} unit vectors in the Hilbert space. (We thank the reviewer for pointing out this subtlety.) We note to expert readers that in situations involving quantum gauge theories, superselection rules, etc., it is not obvious that all unit vectors in the Hilbert space do indeed necessarily represent physical states. But, as discussed in Section~\ref{Formulations}, we do not focus on these more complicated situations in this article.} But then it follows by linearity (axiom 4A') that $\hat{U}$ must preserve the norm of \emph{all} vectors in the Hilbert space: if $|\phi\rangle$ is any nonzero vector, then
\begin{align*}
\left( \frac{\langle \phi |}{\norm{\phi}} \right) \hat{U}^\dagger \hat{U} \left( \frac{| \phi \rangle}{\norm{\phi}} \right) &= 1 \\
 \langle \phi | \hat{U}^\dagger \hat{U} | \phi \rangle &= \norm{\phi}^2 \\
\norm{\hat{U} | \phi \rangle} &= \norm{\phi}.
\end{align*}

Next, we use the result that $\hat{U}$ preserves norms to show that it in fact preserves \emph{all} inner products. The proof is almost identical to the derivation of the \emph{polarization identity} for a complex inner product space, which uses the norm of an arbitrary vector in the space to derive the form of the space's inner product \cite{Halmos}. Consider the generic vector sum $|\alpha \rangle + |\beta\rangle$:
\[
\left( \langle \alpha | + \langle \beta | \right) \hat{U}^\dagger \hat{U} \left( | \alpha \rangle + | \beta \rangle \right) = \left( \langle \alpha | + \langle \beta | \right) \left( | \alpha \rangle + | \beta \rangle \right).
\]
Expanding out the sums, using $\langle \alpha | \hat{U}^\dagger \hat{U} | \alpha \rangle = \langle \alpha | \alpha \rangle$ and $\langle \beta | \hat{U}^\dagger \hat{U} | \beta \rangle = \langle \beta | \beta \rangle$, and simplifying yields
\[
\Re \left[ \langle \alpha | \hat{U}^\dagger \hat{U} | \beta \rangle \right] = \Re \left[ \langle \alpha | \beta \rangle \right].
\]
Similar manipulations starting from the complex linear combination $|\alpha\rangle + i | \beta \rangle$ give that $\Im \left[ \langle \alpha | \hat{U}^\dagger \hat{U} | \beta \rangle \right] = \Im \left[ \langle \alpha | \beta \rangle \right]$, so
\[
\langle \alpha | \hat{U}^\dagger \hat{U} | \beta \rangle = \langle \alpha | \beta \rangle.
\]
Since this identity holds for all vectors $|\alpha\rangle$ and $|\beta\rangle$ in the Hilbert space, we have that $\hat{U}^\dagger \hat{U} = \hat{I}$. For a finite-dimensional Hilbert space, this implies that $\hat{U}$ is unitary.
\end{proof}

So we can weaken axiom 4A to axiom 4A' without changing the resulting theory: formulations A and A' are equivalent.

But the analogous proposition does \emph{not} hold for formulation B. Suppose we considered weakening axiom 4B to an axiom 4B' that is the same as 4A'; that is, we weaken the postulate that the time-evolution operator $\hat{U}$ is unitary to merely require $\hat{U}$ to be linear. Axiom 1B requires that state vectors be nonzero, so the time-evolution operator $\hat{U}$ cannot map a nonzero vector to $0$, so $\hat{U}$ must be invertible (for a finite-dimensional Hilbert space).\footnote{We again make the physical assumption that time evolution must map a valid physical state to a valid physical state.

In full generality, a linear operator $\hat{U}$ does not map any nonzero vector to $0$ iff its kernel is trivial iff $\hat{U}$ is injective. Physically, this means that two different initial states cannot evolve to the same final state, so time evolution must be reversible. For a finite-dimensional vector space, this implies that $\hat{U}$ must be invertible. But for an infinite-dimensional vector space, an injective linear operator might not be surjective, and therefore not invertible. There can be states that cannot be ``reached'' via time evolution from \emph{any} previous state.

Returning to the finite-dimensional case: in this generalized theory, the set of possible time-evolution operators is expanded from the $N^2$-dimensional unitary Lie group $\mathrm{U}(N)$ to the larger $2N^2$-dimensional Lie group $\mathrm{GL}(N, \mathbb{C})$, where $N$ is the dimension of the Hilbert space.} But other than that, its form is not logically constrained by the generalized axioms 1B-4B'.

The generalized axiom set B' allows a state vector's norm to change over time. Does this possibility have physically observable consequences? That is, can axiom set B' produce measurement probability distributions $\{P(\lambda)\}$ that are not possible within the standard axiom set B? The answer is an emphatic yes.

Ref.~\cite{Aaronson} briefly discusses the physical theory (which it refers to as \emph{global manual normalization}\footnote{Strictly speaking, Ref.~\cite{Aaronson} only refers to this theory as ``manual normalization'', but it contrasts this theory with another variant that it refers to as ``local manual normalization''. In this paper, we use the term ``global manual normalization'' to distinguish the theory from local manual normalization.}) described by this generalized axiom set B'. The author points out that this generalized theory has physical consequences that are very different from those of standard quantum mechanics. In particular, it allows entanglement to be used for faster-than-light communication!

To see how, consider two physically separated qubits initially in the entangled Bell state $|\psi_i\rangle = |00\rangle + |11\rangle$.\footnote{This paragraph (alone in the main text of this article) assumes some background in quantum information theory. See the appendix for some additional details that we gloss over in this discussion.} The reduced density matrix for Bob's qubit (listed second) is the maximally mixed state $\hat{\rho}_B = \frac{1}{2} \hat{I}$, and Bob initially has an equal probability $1/2$ of measuring his qubit to have either value 0 or 1. If Alice wants to transmit a $0$ bit to Bob, then she can apply the non-unitary qubit gate
\[
\left(
\begin{array}{cc}
1 & 0 \\
0 & \epsilon
\end{array}
\right)
\]
to her qubit, where $0 < \epsilon \ll 1$. The global state becomes $|\psi_f\rangle = |00\rangle + \epsilon |11\rangle$, which has a different norm from $|\psi_i\rangle$. Bob's reduced density matrix is now
\begin{equation} \label{Bob}
\hat{\rho}_B = \left(
\begin{array}{cc}
1 & 0 \\
0 & \epsilon^2
\end{array}
\right),
\end{equation}
and Bob's new normalized measurement probabilities are $\{P(0) = 1/(1+\epsilon^2),\ P(1) = \epsilon^2/(1+\epsilon^2)\}$, so Bob is almost guaranteed to measure his qubit to have the transmitted value $0$.\footnote{Interestingly, Alice still cannot use entanglement to communicate completely deterministically within this protocol, because $\epsilon = 0$ would make $U$ singular and annihilate the $|1\rangle_A$ state -- although she can make the probability of transmission error arbitrarily small.} Similarly, Alice could have chosen to apply a different non-unitary gate to her qubit in order to transmit the bit $1$. By contrast, the no-communication theorem gives that if Alice can only apply \emph{unitary} operators to her qubit, as in axiom 4B, then she cannot use entanglement to change Bob's reduced density matrix at all, nor any of his measurement probabilities.

Ref.~\cite{Aaronson} also gives a (much more complicated) proof that a hypothetical quantum computer that operated under the generalized axioms B' would be able to solve all problems in the complexity class PP in polynomial time. The complexity class PP is believed to contain a much larger and more difficult set of problems than does the complexity class BQP of problems solvable by a standard quantum computer in polynomial time.\footnote{PP is also believed to contain a larger and more difficult set of problems than the more famous complexity class NP.}

Is it possible to formulate this non-standard physical theory (global manual normalization) in a way that ``looks similar'' to the more familiar formulation A? We have already shown that this theory is not fully compatible with formulation A (which is simply standard QM), but we can formulate global manual normalization in a way that is compatible with axioms 1A-3A at the cost of slightly modifying axiom 4A. In this reformulation, we are not allowed to wait to normalize until a measurement is made; postulate 1A requires that the state vector be normalized at all times. So if a time-evolution process looks like $|\psi\rangle \to \hat{U} |\psi\rangle$ within formulation B (where $\hat{U}$ is not necessarily unitary), then in order to respect axiom 1A in the other formulation, after applying $\hat{U}$ we must additionally ``manually'' rescale the output back to a unit vector. This composed time-evolution map -- the modification of axiom 4A that describes the theory of global manual normalizaion -- takes the form 
\[
|\psi_i \rangle \to |\psi_f \rangle = \frac{1}{\sqrt{ \langle \psi_i | \hat{U}^\dag \hat{U} | \psi_i \rangle}} \hat{U} |\psi_i \rangle,
\]
which is nonlinear because the scalar prefactor depends on the input state $|\psi_i\rangle$ (unless $\hat{U}^\dag \hat{U}$ is proportional to the identity). This perspective makes it clear why global manual normalization leads to very different phenomenology than standard QM does.

Interestingly, global manual normalization \emph{does} reproduce standard QM if the time-evolution operator is \emph{proportional} to a unitary operator, even if the (nonzero) proportionality constant does not equal $1$. Such an operator can be thought of as uniformly dilating the entire Hilbert space. This suggests that the fundamental characteristic of time evolution in QM may be best thought of not as preserving norms per se, but instead as preserving \emph{relative} norms or angles between states.

Of course, we could also consider other ``intermediately strong'' generalizations of axiom 4B, in which we allow \emph{some} non-unitary time-evolution operators, but not the full set of invertible operators. Such theories could perhaps be made compatible with existing experimental results, but they would probably need to be quite convoluted and unnatural.

\section{Conclusion}

This article attempted to address the question of whether the basic mathematical rules of probability alone require that time evolution in quantum mechanics be unitary. The answer turns out to be rather subtle.

We developed two physically equivalent versions of the basic axioms of QM, which each postulate that time evolution is unitary. But while these two formulations are physically equivalent, they naturally \emph{generalize} in different ways to two theories that turn out to be physically distinct. In particular, in one formulation (axiom set A), the postulate of unitary time evolution is indeed unnecessary; the unitarity of time evolution follows logically from the assumption of \emph{linearity} only (and the other axioms). But in the other formulation (axiom set B), the postulate of unitary time evolution is essential. If we weaken that axiom to only postulate linear time evolution, then we end up with a new physical theory that is completely logically self-consistent, but which makes very different experimental predictions than standard QM does.

Of course, all the experimental evidence collected so far supports the hypothesis that time evolution is indeed unitary and is given by the Schr\"{o}dinger equation. But, contrary to what is sometimes loosely implied, this hypothesis of unitarity is experimentally falsifiable, and is not merely a tautological claim that follows from the basic rules of probability and from \emph{every} formulation of the other axioms of QM.

\backmatter

\bmhead{Acknowledgements}

The author thanks the anonymous reviewer for a careful review and very thoughtful feedback.

\section*{Declarations}

The author did not receive funding from any sources to prepare this manuscript and has no competing interests to declare. No data was produced during this research.

\begin{appendices}

\section*{Appendix}

In this appendix, we briefly extend the discussion in the main text to discuss mixed states, multipartite systems, and reduced density matrices. (Unlike in the main text, in this appendix we assume that the reader is familiar with these more advanced concepts in quantum mechanics.) This is a much more complicated subject than the ``QM lite'' covered in the main text, so we will not go into depth but will only briefly touch on how this situation differs from the simpler situation considered in the main text.

In textbook QM, mixed states are postulated to be represented by a \emph{density operator} $\hat{\rho}$: a positive-semidefinite (trace-class) operator on the Hilbert space that is typically defined to be normalized to have trace $1$. This postulate generalizes formulation A discussed in the main text to the case of mixed states. Any physical observable is represented by a self-adjoint operator $\hat{A}$, and the expectation value of the observable for the state $\hat{\rho}$ is postulated to equal $\langle \hat{A} \rangle = \Tr(\hat{\rho} \hat{A})$. This postulate generalizes the postulate 3A discussed in the main text.

Just as in the case of pure states, we can decide whether to formulate the normalization requirements for mixed states as applying either before or at the point of measurement. Instead of the standard normalization conventions stated above (the generalization of formulation A), we can instead generalize formulation B to a set of postulates in which the density operator is trace-class and positive semidefinite, but its trace is unconstrained (within the definition), and the expectation value of an observable $A$ is postulated to equal $\langle \hat{A} \rangle = \Tr(\hat{\rho} \hat{A})/\Tr(\hat{\rho})$.

Just as discussed in subsection~\ref{equiv}, this alternate formulation is completely physically equivalent to the standard textbook formulation; we are simply ``moving'' the mathematical requirement that probabilities sum to $1$ from the normalization of $\hat{\rho}$ to the formula for extracting experimental probabilities.

But a new wrinkle arises in the case of mixed states. A mixed state $\hat{\rho}$ on a Hilbert space $\mathcal{H}_A$ is often (either implicitly or explicitly) assumed to be a reduced density matrix $\Tr_B(|\psi \rangle \langle \psi|)$ that results from taking the partial trace of a pure state\footnote{For conciseness, we will use the term ``pure state'' to refer to either (a) the abstract physical state itself, (b) the \emph{vector} representation $|\psi\rangle$, or (c) the rank-$1$ \emph{operator} representation $|\psi\rangle \langle \psi |$, depending on the context.} $|\psi\rangle \langle \psi |$ in a larger Hilbert space $\mathcal{H}_A \otimes \mathcal{H}_B$. Sometimes, the ``auxiliary'' Hilbert space $\mathcal{H}_B$ is considered to represent the physical environment around the system, whose detailed state is not of experimental interest; other times, $\mathcal{H}_B$ is simply thought of as a formal Hilbert space used to mathematically ``purify'' the mixed state $\hat{\rho}$ to a formal pure state in a larger Hilbert space, which can sometimes be a convenient mathematical technique for certain theoretical analyses. Under the relation $\hat{\rho} = \Tr_B(|\psi \rangle \langle \psi|)$, the density operator is normalized (i.e. $\Tr(\hat{\rho}) = 1$) iff the ``larger'' pure state is normalized (i.e. $\norm{\psi} = 1$).

In the context of reduced density matrices, there are \emph{three} natural choices of where to impose the normalization requirement: ``before'', ``while'', or ``after'' taking the partial trace. More precisely, we have three choices of physically equivalent sets of formulas:
\begin{enumerate}
\item As in the standard textbook formulation (the equivalent of formulation A in the main text), we can require that the ``larger'' pure state $|\psi\rangle$ be normalized. In this case, the relation $\hat{\rho} = \Tr_B(|\psi \rangle \langle \psi|)$ implies that the reduced density operator $\hat{\rho}$ is normalized as well.
\item We can allow the ``larger'' pure state to have arbitrary norm, but modify the relation between the pure state and the reduced density operator to \[
\hat{\rho} = \frac{\Tr_B(|\psi \rangle \langle \psi|)}{\norm{\psi}^2}.
\]
In this case we still automatically have $\Tr(\hat{\rho}) = 1$, so we can still consistently take the expectation value of an observable $\hat{A}$ on $\mathcal{H}_A$ to equal $\langle \hat{A} \rangle = \Tr(\hat{\rho} \hat{A})$.
\item We can allow the ``larger'' pure state to have arbitrary norm and keep the usual relation $\hat{\rho} = \Tr_B(|\psi \rangle \langle \psi|)$ between the pure state $|\psi\rangle$ and the reduced density operator $\hat{\rho}$. In this case, the reduced density operator will generically not be normalized ($\Tr(\hat{\rho}) \neq 1$). In order to be mathematically well-defined, we need to impose the normalization requirement at the point of taking expectation values of observables for the reduced system: similarly to the formulation B discussed in the main text, we must define the expectation value of the obervable $A$ on $\mathcal{H}_A$ to equal $\langle \hat{A} \rangle = \Tr(\hat{\rho} \hat{A})/\Tr(\hat{\rho})$. 
\end{enumerate}
In some sense, we can think of the new case \#2 as being ``intermediate'' between formulations A and B in the main text.

As long as the \emph{global} pure state $|\psi\rangle \langle \psi|$ evolves unitarily, all three of these formulations are physically equivalent to standard QM. But we see that the situation gets rather complicated if we generalize standard QM to allow for the global pure state $|\psi\rangle$ to evolve according to an arbitrary linear operator $\hat{U}$; in that case, these three formulations might generalize in several different inequivalent ways.\footnote{For example, the fact that in formulation \#3 the density operator is not required to be normalized opens up the possibility of generalizing the time-evolution map for subsystem A to be completely positive but not trace-preserving.}

In the example with Alice and Bob that we discussed in the main text, we implicitly chose to generalize choice \#3 above: note that after Alice applies her non-unitary operation to her qubit, the global state $|\psi_f\rangle$ of the entangled pair of qubits is no longer normalized, nor is Bob's reduced density matrix~\eqref{Bob}. We see that even in this very simple case where the global time-evolution operator $\hat{U}$ happens to factorize into a tensor product $\hat{U} = \hat{U}_A \otimes \hat{I}_B$ that only acts on subsystem $A$, if $\hat{U}$ is non-unitary then subsystem $B$ can still experience nontrivial dynamics that ``survive'' the partial trace over subsystem $A$.\footnote{See Ref.~\cite{Aaronson} for another (arguably) natural generalization of standard QM (which it calls \emph{local manual normalization}) that applies to the case where the global time-evolution operator factorizes as $\hat{U} = \hat{U}_A \otimes \hat{I}_B$.} This fact is what leads to the violation of the no-signaling theorem, which hinges critically on the assumption that the global time-evolution operator is unitary.

\end{appendices}

\bibliography{Unitary}% common bib file

\end{document}